\documentclass[a4paper,11pt]{article}
\pdfoutput=1 

\usepackage{jinstpub} 
\usepackage{algorithm}
\usepackage{algpseudocode}
\usepackage{multirow}
\graphicspath{{figures/}}

\DeclareMathOperator*{\Var}{Var}
\title{Optimal parameters estimation for K-edge subtraction imaging using PixiRad-2/PixieIII photon counting detector on a conventional laboratory X-ray micro-tomograph.}

\author[a, b]{R. Granger}
\author[b]{L. Salvo}
\author[a]{S. Rolland du Roscoat}
\author[b]{P. Lhuissier}

\affiliation[a]{Univ. Grenoble Alpes, CNRS, Grenoble INP\footnotemark , 3SR, UMR 5521, F-38000 Grenoble, France}

\affiliation[b]{Univ. Grenoble Alpes, CNRS, Grenoble INP\footnotemark[\value{footnote}], SIMAP, F-38000 Grenoble, France \footnotetext[\value{footnote}]{Institute of Engineering Univ. Grenoble Alpes}}

\emailAdd{remi.granger@univ-grenoble-alpes.fr}
\emailAdd{pierre.lhuissier@simap.grenoble-inp.fr}

\abstract{Photon Counting Detectors (PCDs) open new opportunities in X-ray imaging. Pixie III is a PCD using simultaneously two energy thresholds. This enables to acquire images using two distinct energy bins in a single exposure and allows to perform K-Edge Subtraction (KES) imaging with laboratory sources. In that context, one has however to deal with an energy bin optimization: narrow energy bins lead to high KES signal at the expense of higher noise, while wider energy bins lead to poor KES signal but better statistics. This work presents a model that aims at finding the optimal energy thresholds and source voltage in order to retrieve the best Contrast to Noise Ratio (CNR) for a given sample. The model also optimizes the parameters for conventional absorption modality and compares both modalities. 
Since the input flux and the energy difference between the thresholds influence image noise, this is included in the model using phenomenological laws. The model is then compared to empirical optimization by experimental screening of the parameters using model materials composed of barium, iodine and water. Finally, it is explained how to model the influence of sample composition on the predicted CNR values.}


\begin{document}
\maketitle
\flushbottom

\section{Introduction}
\begin{figure}[h]
\centering
\includegraphics[width=0.95\textwidth]{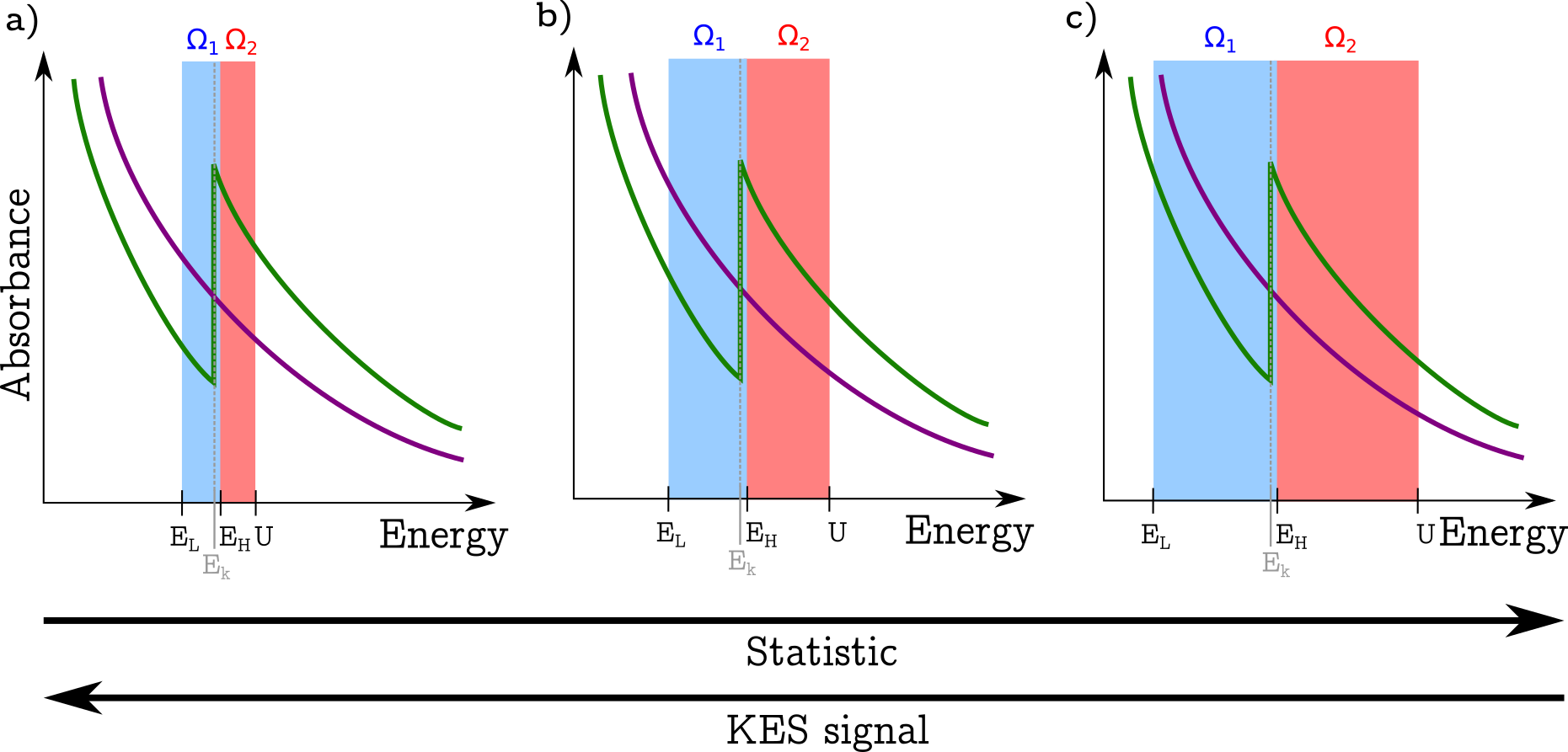}
\caption{\label{fig:scheme} Scheme of the optimization problem of KES imaging with lab source. Green and purple solid curves are absorbance spectra of the two materials. \textbf{a)} Taking narrow energy bins: good signal, low statistic, \textbf{b)} large bins: good statistics, low signal, \textbf{c)} optimal energy bins.}
\end{figure}

X-ray tomography is now daily used in various research domains such as material science, medical and biological science, and more and more in industries (see e.g. \cite{maire2014}).
The conventional and most employed technique is based on absorption contrast, discriminating material by their difference of absorption at a given energy (monochromatic beam) or by their difference of integrated absorption over an energy range (polychromatic beam).
However, it may happen that this contrast is not large enough to differentiate the materials. A possibility for enhancing contrast in that case is to exploit phase contrast techniques  \cite{nugent1996,paganin2006,myers2007}. However these techniques are not easy to handle in laboratory tomographs. Another possibility is to increase contrast from variation in absorption coefficient measured at two different energies (or over two distinct energy ranges).  This is particularly efficient when the K-edge of one of the materials, that corresponds to a strong variation of absorption coefficient,  lies between the two spectral measurements points. Thus, this approach is named K-edge Subtraction (KES) imaging \cite{thomlinson2018}. The classical example of such an application is the use of contrast agent as iodine or barium for biological and medical applications.
Historically, this technique was developed with quasi-monochromatic condition using a laboratory source, with secondary source fluorescence peaks \cite{jacobson1953, rutt1983} or additional monochromator \cite{zhong1997}. This permits measurements to be performed right below and right above the K-edge. However, these approaches were restricted to specific chemical elements and limited to low flux.  Practically, this technique was mainly used at synchrotron \cite{rubenstein1984, elleaume2002} as the beam offer high monochromaticity and high flux. However beam time access is not easy to obtain.
Recent development of new technologies such as Compact  Light Source allows overcoming this difficulty and was used to perform KES imaging  for medical application\cite{kulpe2018}. This kind of sources definitely offers promising perspectives but their access and use are still more complicated than laboratory sources.
Another strategy to develop KES imaging relies upon the recent development of photon counting detectors, which present energy-discriminating capabilities \cite{ballabriga2016}.
One of them is the PixieIII photon counting detector \citep{bellazzini2015}. It has the capacity  to count incoming photons over two distinct energy ranges simultaneously, allowing to acquire two images in a single exposure time when illuminated with a polychromatic beam.
However, in that case, as the absorption coefficient measurements corresponds to mean over energy ranges, and not at specific energies, one has to deal with an optimization problem. 
This problem is depicted in figure \ref{fig:scheme}.
Consider two  absorbance spectra of two distinct materials, one with K-edge at energy $E_K$ and another one without K-edge represented by the green and purple solid curves, respectively.
The PCD detector thresholds $E_L$ and $E_H$ and the source voltage U defines the energy range $\Omega_1$ and $\Omega_2$ over which the two spectral images are formed (represented by the blue and red shaded areas, respectively). 
In order to get the strongest KES signal, one may be tempted to use narrow energy ranges on each side of the K-edge energy as sketched by figure \ref{fig:scheme}.a).
However, since laboratory sources offer limited photon flux, this would result in low statistic images. 
Conversely, in order to get lower noise, one may choose large energy bins as in figure \ref{fig:scheme}.c). In this case, the KES signal is low since the averaging over large energy bin smooths the K-edge.  So the optimal tuning of the energy ranges to perform the K-edge imaging is a trade-off between a high KES signal and a good statistic in the images (fig. \ref{fig:scheme}.b)) 

To address this problem, He et al. \cite{he2012, he2013} presented a theoretical approach to predict the optimal width of the energy range. 
Their model is based on optimization of contrast to noise ratio of the projection images. The noise  is Poisson noise from the beam, and the contrast signal being obtained from Beer-Lambert law through the sample.
Their results assumed an equal width of the two energy bins. Later, Meng et al. \cite{meng2016} used the same approach, the optimization being based on the Contrast to Noise Ratio (CNR) of the reconstructed images, and showed that bin width are similar when optimizing CNR on projections or reconstructed images.
While this approach tackles the optimization problem for the general case, detector-specific features like detector noise or spectral resolution are not taken into account. 
For PixieIII, Brun et al. \cite{brun2020} presented results of empirical optimization of one of the two thresholds on a test phantom made of iodine, water and barium. The objective of this work is to present an optimization model that includes both theoretical modelling of general behaviour of detector and empirical detector-specific characteristics (noise, spectral resolution, counter depth) in order to refine predictions when using PixieIII detector. Also, the model aims at optimizing source voltage and current, both energy thresholds, and predicting whether it is better to use KES imaging or conventional absorption imaging for the considered sample. 
This paper is structured as follows. The first section presents the main equations of the model. Then, the second section is dedicated to
the materials and methods fincluding characterization of the energy resolution in terms of absorbance spectra, noise parametrization, model resolution and validation and application example. The third section presents the results of these studies which are discussed in the fourth section.

\section{Model\label{sec:expressions}}
\subsection{Generality}

The objective of the model is to help the operator to choose configuration parameters based on the sample he wants to image and the objectives of the acquisition.
Given the materials $X$ and $Y$ to be distinguished in the image, with respective densities $\rho_X$ and $\rho_Y$, the model predicts the source voltage $U$, current $i$,  the two detector energy thresholds ($E_L$ and $E_H$) and the modality to use (absorption or KES imaging) in order to achieve the highest Contrast-to-Noise Ratio (CNR) on the obtained radiographs. Actually, the current $i$ and the exposure time have equivalent influence on the variable of interest, namely it acts on the number of incoming photons during exposure. For sake of simplicity the reasoning is done with current $i$ , with a fixed exposure time. But the current variable could then be thought as the product of current times exposure time or
exposure while keeping current fixed without modifying the conclusions. The objective of the model is represented by the function $f$ in the equation \ref{eq:objective}.
\begin{equation}
\label{eq:objective}
    (U, E_L, E_H, i, MOD) = f(X,Y, \rho_X, \rho_Y)
\end{equation}
In the context of computed tomography, \cite{meng2016} showed that optimizing the CNR for radiographs is close to optimizing the CNR on reconstructed images. Thus, we focused in predicting CNR on radiographs,  to perform optimization of 5 parameters, as shown by relation \ref{eq:objective},  including the two energy thresholds.

The metric to evaluate CNR for a given modality MOD is defined as the difference of mean signal between material $X$ and $Y$ divided by a standard deviation representing noise computed on the image with that modality:
\begin{equation}
CNR_{MOD} = \frac{|\langle MOD_X\rangle - \langle MOD_Y\rangle|}{\sigma_{MOD}}
\end{equation}
Where $MOD_m$ represents the image obtained with the modality MOD (absorption or KES) for  material m, $\langle \cdot \rangle$ stands for mean of pixel values over a representative region of Interest (ROI) and $\sigma_{MOD}$ is the mean of the standard deviations measured for material X and Y.

The absorption image and KES image are combination of raw images. In that section, we derive the expression of absorption contrast $C_{ABS}$ and noise standard deviation $\sigma_{ABS}$ as well as  KES contrast $C_{KES}$ and standard deviation $\sigma_{KES}$ from the mean and the standard deviation of count levels of acquired images.

For both modalities, raw data consists of an image $I_0^{\Omega}$ taken without the sample, and a sample image containing a set $I_X^{\Omega}$ of pixels filled with material $X$  and a set $I_Y^{\Omega}$ of pixels filled with material $Y$. $\Omega$ in these notations refers to the energy range $\Omega_1$ or $\Omega_2$ used for imaging.
The input flux on the detector at a given energy $E$ will be denoted $I_0(E)$. Assuming random noise as unique source of dispersion, the image without sample is flat with a  mean count given by
\begin{equation}
\label{eq:I0}
\overline{I_0^{\Omega}}= \int_{\Omega}I_0(E)dE
\end{equation}for the image with the sample, the mean count is computed using
\begin{equation}
\label{eq:Im}
\overline{I_m^\Omega} = \int_{\Omega} I_0(E)e^{-A_m(E)} dE
\end{equation}for pixels of a sample image filled with material $m \in \{X,Y\}$.  $A_m(E)$ refers to the absorbance of material $m$ and is given by : 
\begin{equation}
\label{eq:absorbance}
    A_m(E) = \frac{4\pi E l_m}{hc}\Im(n_m(E))
\end{equation}
Where $\Im(n_m(E))$ is the imaginary part of the refractive index $n_m(E)$ for a given energy $E$ and a given material $m$ with density $\rho_m$ and thickness $l_m$. $h$ is the Planck constant and $c$ is the speed of light. The refractive index $n_m(E)$ is taken from xraylib database \citep{schoonjans2011}, which gives data with 1 keV spectral resolution.

Since in practice the  flat image is an image acquired with high statistics in order to correct for systematic dispersion of the gray levels, the dispersion of values in flat images will be assumed negligible in comparison with the one of sample images, corresponding to a standard deviation noted $\sigma_m^{\Omega}$,  computed on image $I_m^{\Omega}$.

\subsection{Absorption modality}
Let us start by establishing the expression of the contrast and noise for absorption modality. The absorption image is obtained by dividing the sample image with the flat image.  The absorption image, gathering counts obtained with both registers of Pixie III detector, counting  over $\Omega_1$ and $\Omega_2$ respectively, is given by 
\begin{equation}
ABS_m = \frac{I_m^{\Omega_1}+I_m^{\Omega_2}}{I_0^{\Omega_1}+I_0^{\Omega_2}}
\end{equation}
Note that as Pixie III is a photon counting detector, with energy discriminating thresholds so that images without X-Rays should be (and are in practice in our tests) perfectly dark, there is no need to correct from any dark level  as it is usually done with charge integrating detectors.
The mean is obtained by Taylor expansion \cite{benaroya2005} of a bivariate function for the ratio function, to give :
\begin{equation}
\label{eq:E_abs}
\langle ABS_m \rangle = \frac{\overline{I_m^{\Omega_1}}+\overline{I_m^{\Omega_2}}}{\overline{I_0^{\Omega_1}}+\overline{I_0^{\Omega_2}}}
\end{equation}
Since $\frac{\Var(I_0^{\Omega_1}+I_0^{\Omega_2})}{\langle I_0^{\Omega_1}+I_0^{\Omega_2}\rangle^2}\ll 1$  by assumption, where $\Var(x)$ denotes the variance of the statistical variable $x$. Similarly, assuming $\frac{\Var(I_0^{\Omega_1} + I_0^{\Omega_2})}{\langle I_0^{\Omega_1} + I_0^{\Omega_2} \rangle} \ll \frac{\Var(I_m^{\Omega_1} + I_m^{\Omega_2})}{\langle I_m^{\Omega_1} + I_m^{\Omega_2}\rangle} $ (low noise in flats compared to sample images), allows expressing the associated variance as:
\begin{equation}
\label{eq:var_abs}
\Var(ABS_m) = \frac{(\sigma_m^{\Omega_1})^2 +(\sigma_m^{\Omega_2})^2  }{(\overline{I_0^{\Omega_1}}+\overline{I_0^{\Omega_2}})^2}
\end{equation}Then,  the CNR for absorption modality is computed
as:
\begin{equation}
\label{eq:cnr_abs}
CNR_{ABS} = \frac{|\langle ABS_X \rangle - \langle ABS_Y \rangle|}{\sigma_{ABS}}
\end{equation}
where $\sigma_{ABS} =0.5 \times (\sqrt{\Var(ABS_X)} +\sqrt{\Var(ABS_Y)})$ is the mean of standard deviation of absorption image computed between pixels filled with materials $X$ and $Y$ respectively. 

\subsection{KES modality}
KES image may be defined as :

\begin{equation}
KES_m = \log  \frac{I_m^{\Omega_2}}{I_0^{\Omega_2}} - \log  \frac{I_m^{\Omega_1}}{I_0^{\Omega_1}}
\end{equation}
Following similar assumptions used to obtain equations \ref{eq:E_abs} and \ref{eq:var_abs}, one obtains:
\begin{equation}
\left \langle \frac{I_m^{\Omega_i}}{I_0^{\Omega_i}} \right \rangle = \frac{\overline{I_m^{\Omega_i}}}{\overline{I_0^{\Omega_i}}}
\end{equation}and 
\begin{equation}
\Var\left(  \frac{I_m^{\Omega_i}}{I_0^{\Omega_i}} \right) = \frac{(\sigma_m^{\Omega_i})^2}{\overline{I_0^{\Omega_i}}^2}
\end{equation}
where i=1 or 2. Then, using first order Taylor expansion of the $\log$ function \cite{benaroya2005}, one may estimate:
\begin{equation}
\left \langle \log  \frac{I_m^{\Omega_i}}{I_0^{\Omega_i}}\right \rangle = \log  \frac{\overline{I_m^{\Omega_i}}}{\overline{I_0^{\Omega_i}}} - \frac{\Var\left(  \frac{I_m^{\Omega_i}}{I_0^{\Omega_i}} \right)}{2\cdot\left(\frac{\overline{I_m^{\Omega_i}}}{\overline{I_0^{\Omega_i}}}\right)^2}
\end{equation}
and 
\begin{equation}
\label{eq:var_ln}
\Var\left (\log  \frac{I_m^{\Omega_i}}{I_0^{\Omega_i}}\right ) = \frac{\Var\left(  \frac{I_m^{\Omega_i}}{I_0^{\Omega_i}} \right)}{\left(\frac{\overline{I_m^{\Omega_i}}}{\overline{I_0^{\Omega_i}}}\right)^2} - \frac{\Var\left(  \frac{I_m^{\Omega_i}}{I_0^{\Omega_i}} \right)^2}{4\cdot\left(\frac{\overline{I_m^{\Omega_i}}}{\overline{I_0^{\Omega_i}}}\right)^4}
\end{equation}
this lead to
\begin{equation}
\langle KES_m \rangle= \left \langle \log  \frac{I_m^{\Omega_2}}{I_0^{\Omega_2}}\right \rangle -  \left \langle \log  \frac{I_m^{\Omega_1}}{I_0^{\Omega_1}}\right \rangle
\end{equation}
and
\begin{equation}
\label{eq:var_kes}
 \Var (KES_m) = \Var\left (\log  \frac{I_m^{\Omega_2}}{I_0^{\Omega_2}}\right ) + \Var\left (\log  \frac{I_m^{\Omega_1}}{I_0^{\Omega_1}}\right )
\end{equation}
Thus, the CNR for KES modality may be expressed as:
\begin{equation}
\label{eq:cnr_kes}
CNR_{KES} = \frac{|\langle KES_X \rangle - \langle KES_Y \rangle| }{\sigma_{KES} }
\end{equation}
where $\sigma_{KES} = 0.5 \cdot (\sqrt{ \Var (KES_X) }+\sqrt{ \Var (KES_Y) })$.

Equation \ref{eq:cnr_abs} and \ref{eq:cnr_kes} give expression of CNR computed from the input spectrum $I_0(E)$, the gray level $\overline{I_m^{\Omega_i}}$ for both material and noise $\sigma_m^{\Omega_i}$ on raw data. These quantities, specific to the hardware used, can be seen as hardware configuration parameters of the model. The method we used to obtain them will be described in section \ref{sec:materials}.

\section{Materials and Methods\label{sec:materials}}

In this section, we first present the hardware used in our case and the test sample.
Then, we describe methods to model or to obtain the empirical parametrization of  $I_0(E)$, $\overline{I_m^{\Omega_i}}$ and $\sigma_m^{\Omega_i}$. Last, we present the optimization scheme, the method for validating the model, and finally an application example.

\subsection{The hardware: source and detector\label{sec:detector}}

For every acquisitions, we used a microfocus reflection source (HAMAMATSU L12161-07, Hamamatsu, Japan) used with a tungsten anode and operated with an acceleration voltage of 50 kV.

The detector is in a PixiRad2-PixieIII, a photon counting detector with energy discriminating capabilities. 
The detector has 2 PixieIII module tiled side-by-side based on a 750 µm thick CdTe crystal bonded to a processing ASIC. This offers a grid of 1024x402 squared pixels of 62 µm. 
The detector allows to select two thresholds -- that will be named low threshold, at energy $E_L$ and high threshold at energy $E_H$ -- associated with two 15 bit counters. Such a configuration enables to acquire 2 images, using two distinct energy bins simultaneously. 
The so called low-energy counter counts photons with energy in range $\Omega_1 = [E_L,E_H]$ while the high energy counter counts photons with energy in range $\Omega_2 = [E_H, \infty )$. 
The detector has the specificity to offer 3 levels of correction of charge sharing at the hardware level. Throughout this article, the detector was operated in NPISUM mode, which corresponds to the finest level of correction of charge sharing. This mode corrects both the number of counts, and the energy of the interaction events (\cite{bellazzini2015,ditrapani2018,ditrapani2020}).

\subsection{Test sample\label{sec:phantom}}

\begin{figure}[h]
\centering
\includegraphics[width=0.5\textwidth]{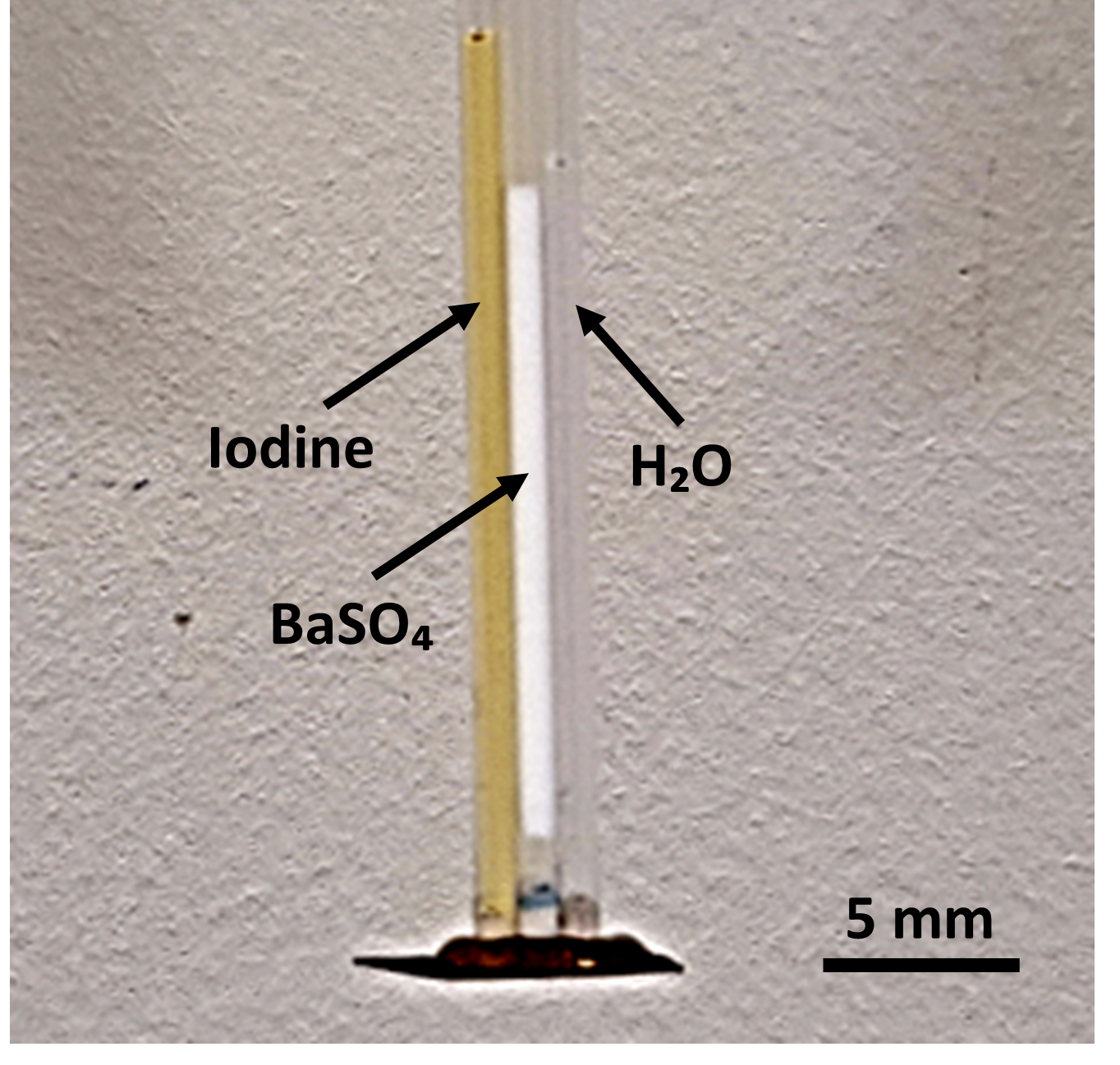}
\caption{Test phantom made with three glass tubes of 1 mm of inner diameter, filled with KI solution, BaSO$_4$ solution and water. \label{fig:phantom}}
\end{figure}

To developand evaluate the performances of the absorption model we used a phantom sample composed of three materials filled in three capillary glass tubes of 1 mm inner diameter. The materials are a solution of 0.3 g/cm$^3$ of KI, a solution of  0.46 g/cm$^3$ of BaSO$_4$ and distilled water for reference (figure \ref{fig:phantom}).

\subsection{Source model\label{sec:source_model}}
Equations \ref{eq:I0} and \ref{eq:Im}, require evaluating $I_0(E)$ over the considered energy ranges. This can be done either using simulated spectrum or  using a spectrometer to obtain an experimental one.
We used SpekPy (\citep{poludniowski2021, bujila2020} to simulate the input spectrum with tungsten anode reflection sources, at desired voltage and current and any source-detector distances.

Alternatively,  we can also obtain an input spectrum by doing a threshold scan with Pixie-III  (i.e., acquiring flat images with increasing threshold value and then differentiating successive images to obtain a spectrum).

\subsection{Absorption model \label{sec:contrast_emmpirical}}
In order to evaluate the mean counts in the sample image $\overline{I_m^{\Omega_i}}$, one needs to compute the absorbance spectra $A_m(E)$ for any energy $E$.
As described in the section \ref{sec:expressions}, this is basically done using xraylib. However, due to the finite resolution of the detector, the measured K-edges are smoother than the tabulated values. In order to take into account that effect, we measured absorbance spectra with the phantom sample (figure \ref{fig:phantom}).  Images with and without the phantom were acquired with the detector using the low register $E_L$ varying from 15 to 50 keV by steps of 1 keV. The mean number of count per pixel for images without the sample ranges from 20 000 for $E_L$=15keV to 500 for $E_L=40$ keV. Then, successive images were differentiated to obtain the number of counts per energy bins of 1 keV width. 
For each energy bins, the image with the sample  was divided by the image without the sample, and the opposite of the logarithm of the result was computed to obtain spectral image series of the absorbance. 
For each of the three tubes, a mean spectrum was computed as the mean value over an ROI consisting of a column of pixel in the centre of the tube.
Finally,  the spectrum obtained from the water tube is subtracted from those obtained from the iodine and barium tubes in order to remove contributions from the glass tube and water. This results in the absorbance spectra named $A_{KI}(E)$ and $A_{BaSO_4}(E)$ of iodine and barium sulfate respectively. 

\subsection{Noise model}
\label{sec:noise_model}

The last part of the model is the prediction of the standard deviation $\sigma_m^{\Omega_i}$ that one can expect on the acquired images.

As mentioned in the introduction, previous papers (\cite{he2012,he2013,meng2016}) focused on modelling Poisson noise from the source, we will now focus on the noise arising from the acquisition procedure as a function of threshold values.
To that purpose, we acquired a set of images without sample, taking all possible combinations of thresholds, between 25 keV and 46 keV by step of 1 keV and using the low energy register, i.e.  counting photon in range $[E_L, E_H]$. For each combination of threshold, 50 images with 1s exposure time were taken. 

Then, we processed the data as follows: for each combination of thresholds, a flat field corrected image was obtained by dividing the first image of the 50 images series by the median of these 50 image. The resulting image is then multiplied by the mean value of the median image. This  permits to reduce gray level variation arising from the inhomogeneity of the beam and pixel to pixel systematic variations. Therefore, the gray level variation is mainly due to noise, so that the image fit the assumptions of the model described in section \ref{sec:expressions}.

Pixels with a zero value in the median image were labelled as defective.
Then, for each combination of thresholds, mean count level $I$  and standard deviation $\sigma$ were computed on a ROI of 1850 pixels (370x5) of the flat-field corrected image, discarding defective pixels. Then, from this dataset, the relations between $I$, $\sigma$ and the two thresholds $E_L$ and $E_H$ were studied and parametrized. Results of this analysis will be found in section \ref{sec:res_noise_model}.

\subsection{Optimization \label{sec:optimization}}
Given the output of the source, the absorption and the noise models, it is possible to evaluate the CNRs for both modalities from configuration parameters using equations \ref{eq:cnr_abs} and \ref{eq:cnr_kes}.  Now, it remains to specify the procedure to find the optimum and the associated constrains.

The procedure is detailed in algorithm \ref{algo}. As a first implementation, the principle is exhaustive, i.e. it consists in evaluating CNRs for both modalities at all points of discrete search ranges $\Omega_U$ (for voltage), $\Omega_{E_L}$ (for $E_L$) and $\Omega_{E_H}$ (for $E_H$). Optionally, during the computation of CNRs, the current is set so that a prescribed count level $I_{max}$ is obtained on the detector when the sample is out of the beam for each set of  voltage and thresholds. 

Then, raw images mean counts $\overline{I_m^{\Omega_i}}$ and $\overline{I_0^{\Omega_i}}$, noise and raw images $\sigma_m^{\Omega_i}$ are estimated. From these variables, CNR for both absorption and KES modality are computed. Once these computations for all combinations of source  voltage and detector thresholds are performed, the best configuration is extracted by searching for the maximum CNR obtained. 

\begin{algorithm}[h]
\caption{\label{algo} Algorithm to optimize $(U,E_L,E_H, \text{mode}, i)$ for maximum  $CNR$}
\begin{algorithmic}
\ForAll {$(U,E_L,E_H) \in \Omega_U \times \Omega_{E_L} \times \Omega_{E_H}$}
\State Compute source spectrum $I_0(E) \forall E \in [0, U]$ for a seed current $i_s$
\State Compute current to reach count level $I_{max}$ on registers:  $i(U,E_L, E_H) = \frac{I_{max}}{max(I_0^{\Omega_1}, I_0^{\Omega_2})} \cdot i_s $
\State Estimate mean gray level on raw images $\overline{I_m^{\Omega_1}}$, $\overline{I_m^{\Omega_2}}$, $\overline{I_0^{\Omega_1}}$, $\overline{I_0^{\Omega_2}}$ using results of section \ref{sec:res_abs_model}
\State Estimate noise on raw images $\sigma_m^{\Omega_1}$ and $\sigma_m^{\Omega_2}$ using results from section  \ref{sec:res_noise_model}.
\State Compute absorption CNR using eq. \ref{eq:cnr_abs}
\State Compute KES CNR using equation \ref{eq:cnr_kes}
\EndFor
\State Find optimum $(U_{opt}, E_{L,opt}, E_{H,opt})$, mode and current $ i(U_{opt}, E_{L,opt}, E_{H,opt})$  for highest CNR.
\end{algorithmic}
\end{algorithm}

\subsection{Validation of the global model \label{sec:validation}}
In order to evaluate the model, we compared predictions to empirical optimization performed on a test sample by evaluating K-edge CNR for all combination of thresholds in given ranges.
The dataset was obtained using the phantom sample described in section \ref{sec:phantom}. This sample allows testing the model on three contrast pairs :   I/BaSO$_4$, water/BaSO$_4$ and water/KI. For that acquisition, the source is operated with 50 kV  acceleration voltage, a target current of 14µA, exposing for 1s at an SDD = 281.12 mm, and SOD = 25.08 mm, thresholds varying from 25 to  45 keV by step of 1keV. 
For each couple of low and high thresholds within that range and for each energy bin, 50 beam images are acquired and 2 images of the sample are taken.
The images were averaged in order to get one beam image and one sample image per couple of thresholds from which KES images were computed. Then, CNR is computed using mean count level  and standard deviation measured over ROIs focusing on each material. The ROIs were made of 3600 pixels (360x10) centred on their corresponding tubes. While the 10 pixels width implies a heterogeneous thickness of tube crossed by the X-rays, it is necessary to get sufficient statistics. In parallel, the simulation is performed using the model presented in the previous section. The results of this validation are presented in section \ref{sec:res_validation} first in terms of the intermediate variable
$I_m^{\Omega_i}$ and $\sigma_m^{\Omega_i}$ and then in terms of CNR and optimal configuration.

\subsection{Application example \label{sec:appli} }

In order to illustrate the application of the model, we show the optimization results with BaSO$_4$ and KI, fixing the density of BaSO$_4$ and varying the density of KI.

The materials whose contrast has to be optimized exhibit a thickness of 1 mm of BaSO$_4$ at 0.46g/cm$^3$, and 1 mm of KI at density in range [0.1g/cm$^3$, 0.5g/cm$^3$].
For each density of KI, algorithm \ref{algo} is applied with $\Omega_U$ = [50,80,100],  $\Omega_{E_L} = \Omega_{E_H} = \llbracket 25\text{ keV}, 45\text{ keV}\rrbracket$  and $ CL = 10 000$.

\section{Results}
\subsection{Absorption Model \label{sec:res_abs_model}}

\begin{figure}[h]
    \centering
    \includegraphics[width=\textwidth]{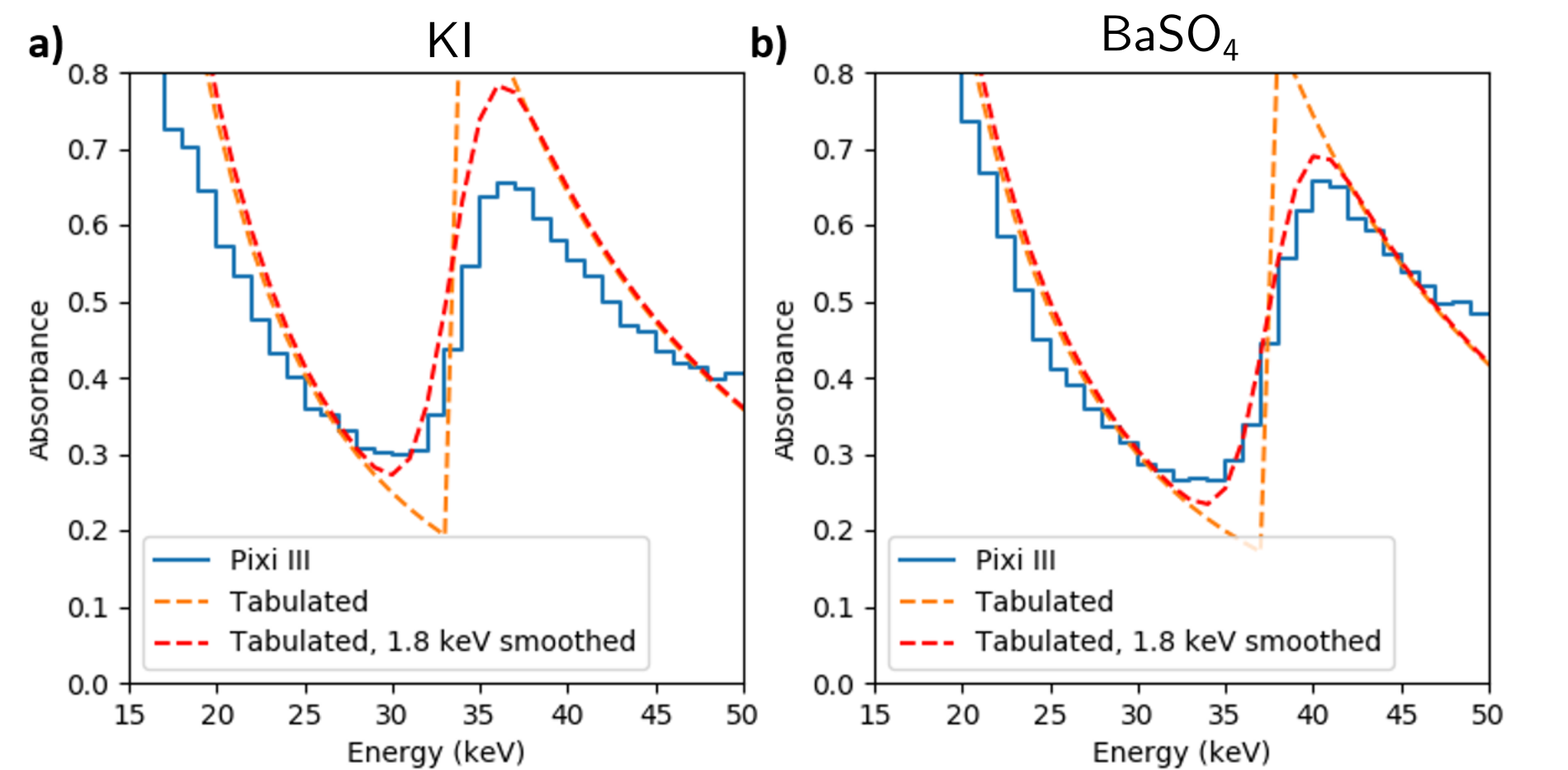}
    \caption{Experimental absorbance spectra for  \textbf{a)} iodine and \textbf{b)} BaSO$_4$ solution measured with PixiRad-2/PixieIII (solid blue curve) compared to computed ones from tabulated value of refractive index (orange dashed line), and computed one with Gaussian kernel of 1.8 keV (red dashed line).}
    \label{fig:spec_res}
\end{figure}
The spectra measured with the detector are the light blue solid curves in figure \ref{fig:spec_res}. The tabulated spectra obtained using xraylib are displayed in orange dashed line. As one can see, the position of the K-edge measured by the detector is correct, as well as the global evolution of the absorbance. However, one needs to account for the spectral resolution of the detector which is necessarily finite and leads to a smoothing of the edges of the absorbance curve. That is why we don't observe a strict vertical edge. This is done by applying a Gaussian smoothing with a 1.8 keV kernel on the curve obtained with tabulated values. This value was chosen in order to reproduce the experimental slope of the edge. The result is represented by the red dashed line in figure \ref{fig:spec_res}.  As one can see, the resulting spectra are really close to the measured ones for both chemical components. In the model, smoothed value of $A_m(E)$ (eq. \ref{eq:absorbance}) are used to compute $\overline{I_m^{\Omega_i}}$ terms.

\subsection{Noise model\label{sec:res_noise_model}}

\begin{figure}[h]
    \centering
    \includegraphics[width=0.8\textwidth]{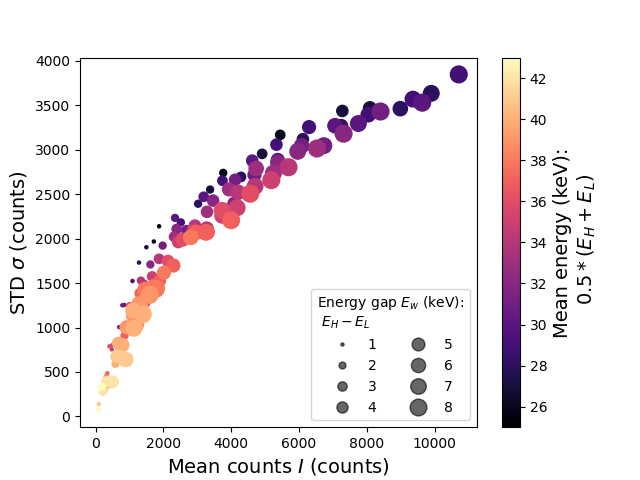}
    \caption{Evolution of standard deviation with mean counts. Colour refers to the mean of the two energy thresholds and size of dots to the difference of the two energy thresholds.}
    \label{fig:std}
\end{figure}
Figure \ref{fig:std} shows the variation of the relation between the mean counts level and the standard deviation when varying the mean energy $E_{mean} = 0.5*(E_L+E_H)$ and the energy gap width $E_w = E_H - E_L$. As one may observe, the relation between the noise and the input flux is dependent on the values of the thresholds.
In order to investigate that dependence, we modelled the measured noise as being the beam intrinsic  noise $\sigma_{intr}$  (proportional to the square root of the input flux)  multiplied by a prefactor  $\Gamma$ that may depend on the thresholds values and possibly the mean counts:
\begin{equation}
\label{eq:prefactor}
\sigma  = \sigma_{intr} \cdot \Gamma (E_{mean}, E_{w}, I) 
\end{equation}
With intrinsic noise  $\sigma_{intr} = \sqrt{I}$ (thus letting the proportionality factor in the prefactor), the prefactor can be rewritten as:
\begin{equation}
\label{eq:prefactor_expr}
\Gamma = \frac{\sigma}{\sqrt{I}}
\end{equation}

\begin{figure}[h]
    \centering
    \includegraphics[width=0.8\textwidth]{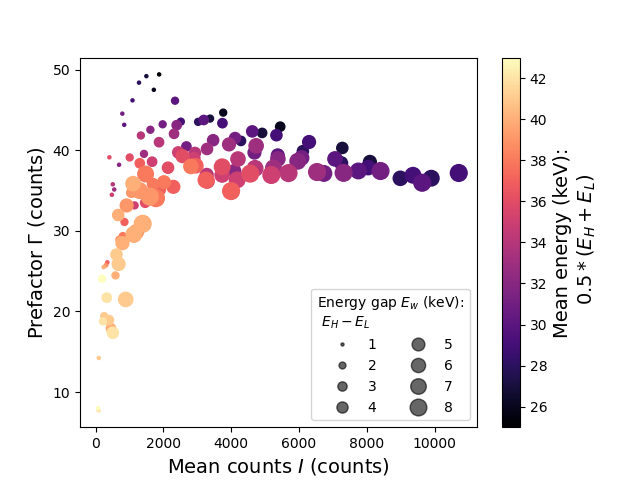}
    \caption{Evolution of the prefactor $\Gamma$  computed with eq. \ref{eq:prefactor_expr} as a function of the mean counts. Colour refers to the mean of the two energy thresholds and size of dots to the difference of the two energy thresholds. }
    \label{fig:prefactor}
\end{figure}

This prefactor is represented in figure \ref{fig:prefactor} as a function of $I$ for eight values of  $E_w$, and for various $E_{mean}$. 
One can see a first dependence of $\sigma$ with $I$ and/or $E_{mean}: $ $\sigma$ increases with $I$ while  $E_{mean}$ decreases.
We assumed that this dependence can be described by $I$ only since the experiment is performed without filter on the source: it is expected that the mean count decreases with the bin mean energy. This point will be discussed in section \ref{sec:discussion}.
Additionally, a  clear dependence can be observed on $E_w$: the points of different size appear to align along different curves. This will also be discussed in section  \ref{sec:discussion}.

\begin{figure}
    \centering
    \includegraphics[width=\textwidth]{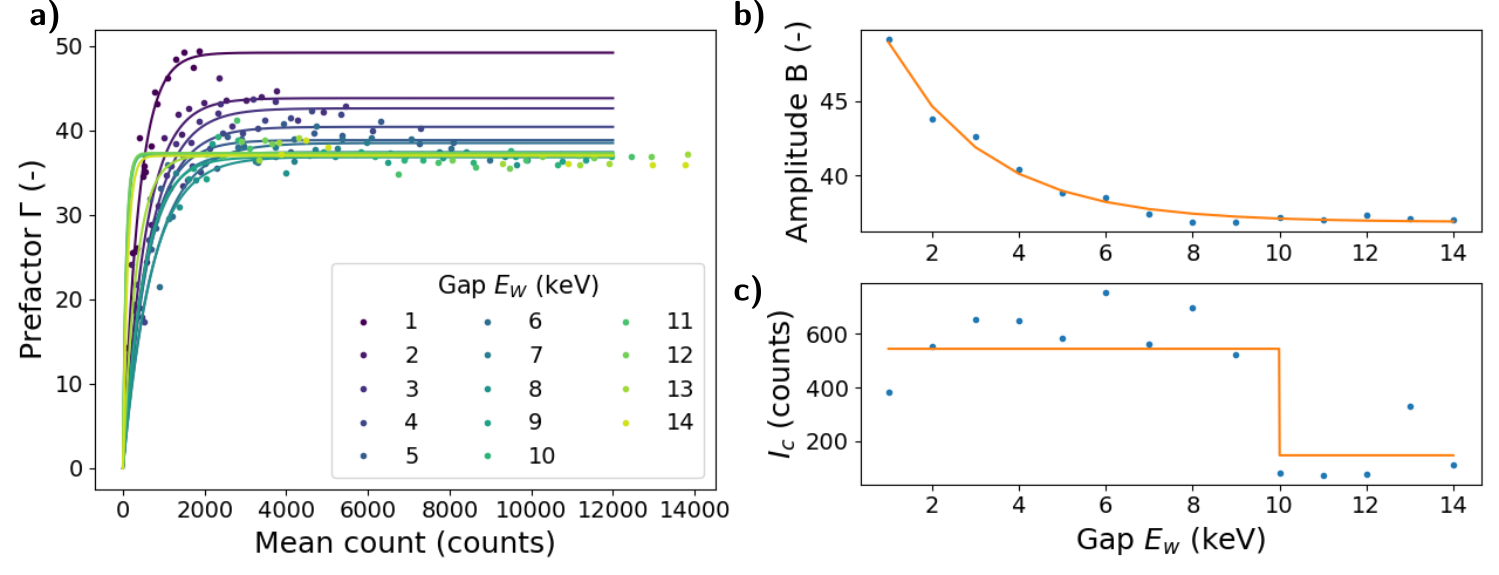}
    \caption{\textbf{a)} Evolution of the prefactor  $\Gamma$ as a function of the mean counts level $I$ for different bins (dots), and fitted model from eq. \ref{eq:fit_shape}. Evolution of \textbf{b)} the fitting parameter amplitude $B$ and \textbf{c)}  characteristic intensity $I_c$ as a function of $E_w$ (dots), and fitted parametrization given in eq \ref{eq:fit} (orange solid line). }
    \label{fig:fit}
\end{figure}
In order to parametrize that dependence, the left panel of figure \ref{fig:fit}a) presents the evolution of the prefactor $\Gamma$ against mean counts $I$ for different fixed $E_w$ (dots). From the evolution, to describe the relation between prefactor $\Gamma$, $E_w$ and $I$, we propose a function of the form:
\begin{equation}
\label{eq:fit_shape}
\Gamma(E_w,I) =  B(E_w)(1-e^{-\frac{I}{Ic(E_w)}})
\end{equation} 
where $B$ and $I_c$ are the fitting parameters (amplitude and characteristic intensity) that depend on the gap width $E_w$.
This shape was fitted on the data experimental data using least square optimization (solid curves on figure 6.a)) and the evolution of the fitting parameters $B$, and $I_c$ against $E_w$ are displayed on figures 6.b) and 6.c).
Evolution of $B$ is monotonically decreasing whereas $I_c$ is approximately increasing up to a given value then does not evolve. Figure 6.a) shows that for largest gap width, the variation of the prefactor is strong where count rate is low. Thus, we suggest parametrizing these evolutions with a decreasing exponential shape for $B$ and a piecewise constant function for $I_c$. Least square fittings of these shapes give:
\begin{equation}
\label{eq:fit}
\begin{split}
    B(I) &= 36.9 +  18.8 e^{-E_w/2.28}\\
      I_c(I) &= 
\begin{cases} 
      544 &\text{ if } E_w\leq 10\text{keV} \\
      147 & \text{else}
   \end{cases}\\
\end{split}
\end{equation}
for $E_w$ in keV.

In summary,  equations \ref{eq:prefactor}, \ref{eq:fit_shape} and \ref{eq:fit}  a model the noise $\sigma$ from the mean count $I$ on detector, and the detector thresholds, or more precisely, the energy gap width $E_w$. 

\begin{figure}[h]
    \centering
    \includegraphics[width=0.56\textwidth]{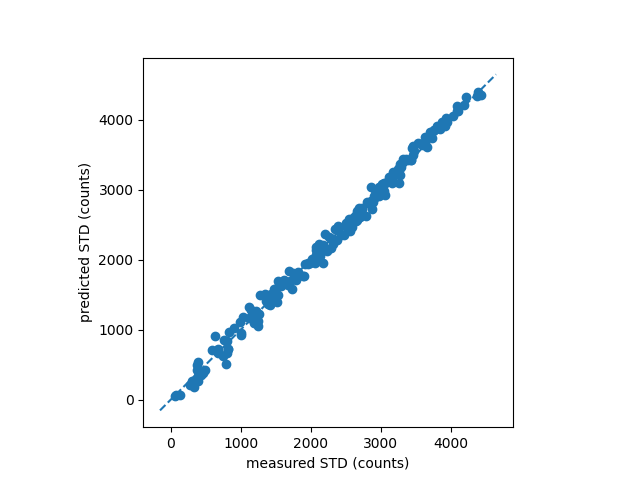}
    \caption{Validation of the model:  predicted standard deviation compared against measured standard deviation from experimental dataset.}
    \label{fig:noise_check}
\end{figure}

We checked the goodness of fit by plotting the estimated standard deviation from $I$ and $E_w$ against the measured one in figure \ref{fig:noise_check}. On this figure, one can see that the points align along the dashed line which the $y=x$ line. It means that the developed parametrization succeeds in explaining the variability of the dataset.

\subsection{Validation  of the global model\label{sec:res_validation}}

\begin{figure}[h]
    \centering
    \includegraphics[width= 0.95\textwidth]{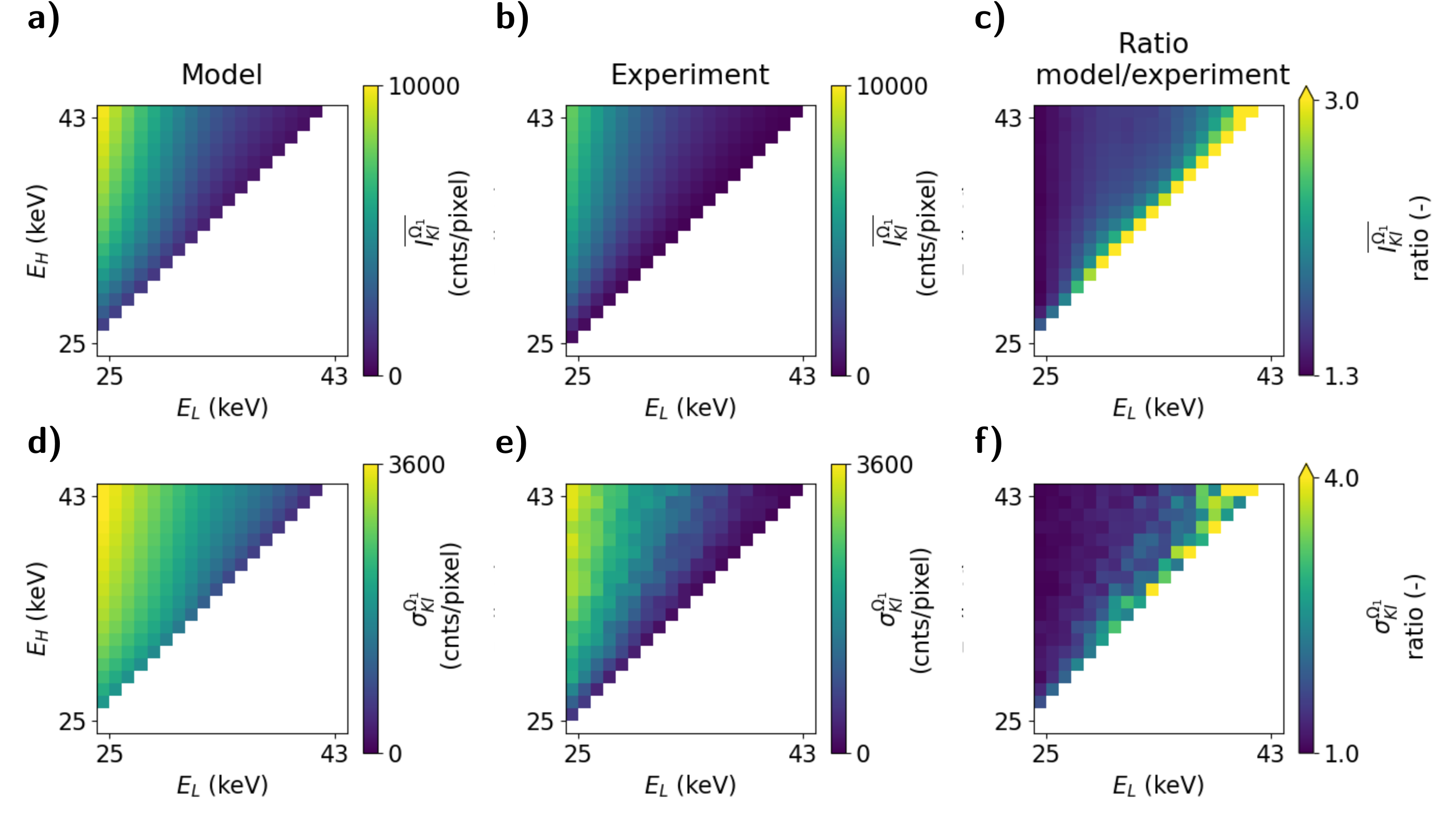}
    \caption{Comparison of intermediate variables as function of the two energy thresholds $E_H$ and $E_L$. \textbf{a), b), c)} $\overline{I_{KI}^{\Omega_1}}$, \textbf{d), e), f)} $\sigma_{KI}^{\Omega_1}$. \textbf{a), d)} simulated quantities, \textbf{b), e)} experimental quantities, \textbf{c), f)} ratio between simulated and experimental.}
    \label{fig:intermediate_var_comp}
\end{figure}

As the model consists in different sub-models, let us first compare some intermediate variables to evaluate these different sub-models.
Figures \ref{fig:intermediate_var_comp} a), b) and c)  compare $\overline{I_{KI}^{\Omega_1}}$, the mean gray level on the ROI focusing on KI tube measured on the low energy register; figures \ref{fig:intermediate_var_comp}  d), e) and f) $\sigma_{KI}^{\Omega_1}$, the standard deviation measured on the same ROI and the low energy register.
We can see that for both variables, simulated values are close to the experimental ones (figure \ref{fig:intermediate_var_comp}c) and f) ). Experimental and simulated maps are qualitatively similar in shape, the values are globally well respected, particularly where  $E_w$ and the number of counts are highest. For $E_w$ getting close to 1, the error increases and the ratio of  simulated values over experimental one  is globally about 5-6.

\begin{figure}[h!]
\centering
\includegraphics[width=0.9\textwidth]{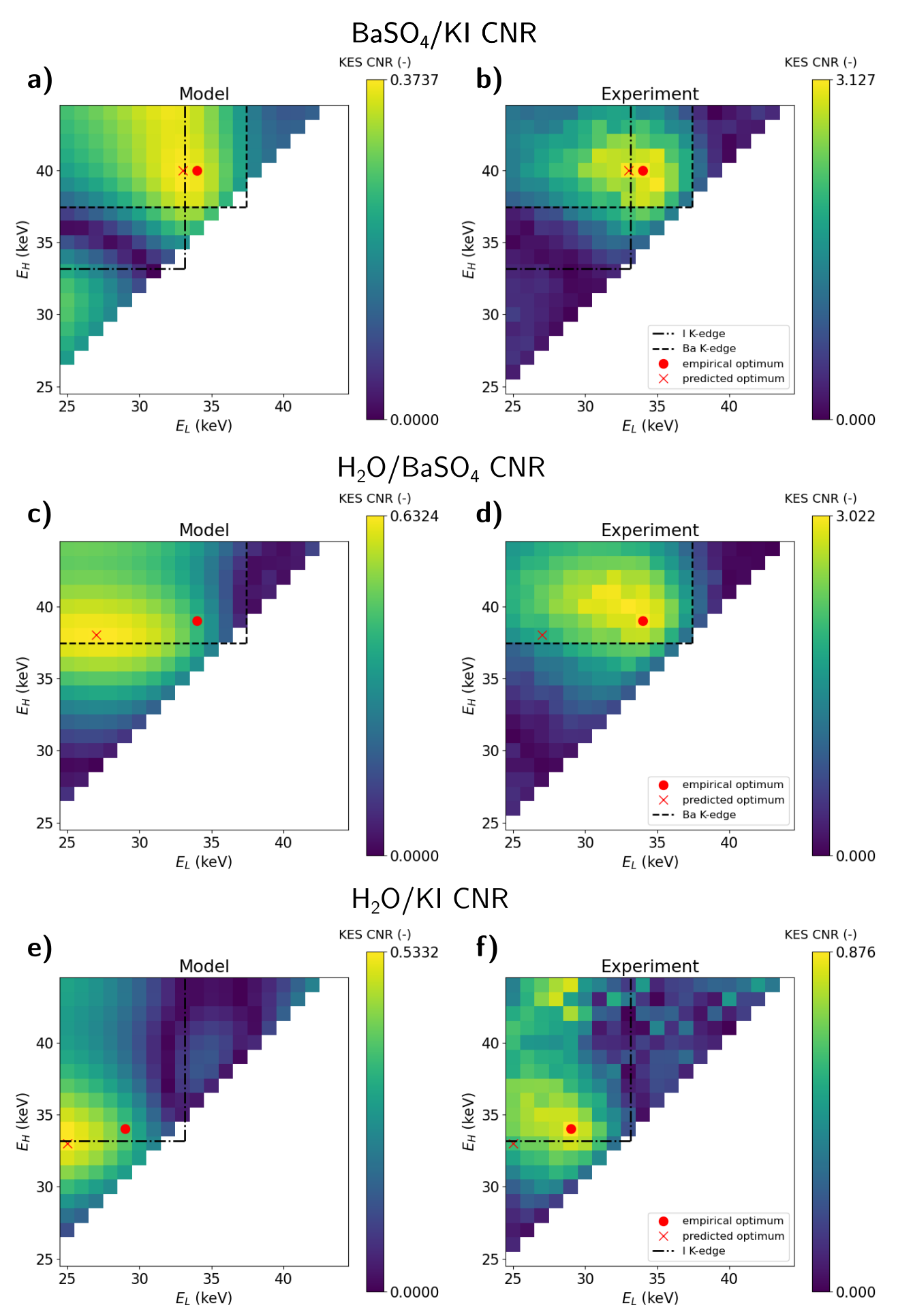}
\caption{Comparison of simulated ( \textbf{a), c), e)} ) and experimental ( \textbf{b), d), f)} ) CNR maps for K-edge modality obtained with  BaSO$_4$ / KI / water test phantom: BaSO$_4$/ KI ( \textbf{a), b)} ), water/ BaSO$_4$ ( \textbf{c), d)} ) and water/KI  ( \textbf{e), f)} ) \label{fig:CNR_maps}}
\end{figure}
Then, we compare CNR maps obtained in KES mode with respect to the two threshold values (see figure \ref{fig:CNR_maps}). The optimal configurations are defined as the maximum of these CNR maps. The empirical optimum is represented by the red dot and the simulated one by the red cross. Additionally, we represent the value of the corresponding K-edges of I by black dotted-dashed lines and of Ba by dashed lines.

Let's begin the comparison with the top row, i.e. for the BaSO$_4$/KI contrast pair (figure \ref{fig:CNR_maps} a) and b) ).
Here again the simulated and experimental maps are similar in shape. From top right to bottom left, we can observe CNR is low for any configuration with both thresholds  above both K-edges; a principal maximum is reached for low thresholds between both K-edges, and high threshold above Ba K-edge. 
A secondary maximum is found at the bottom left part of the map where both thresholds are below both K-edges and a narrow “valley” of minima separates both maxima.

Quantitatively, values for the predicted and measured optima for the three contrast pairs are given in table \ref{tab:optima}.
For the BaSO$_4$/KI pair, the model predicts an optimum at (33 keV, 40 keV)  with a CNR of 0.4 while the empirical optimum is (34 keV, 40 keV) with a CNR of 3.1. 
The secondary maximum for model and experiment is respectively about 0.3 and 0.4.
\begin{table}[h]
\centering
\begin{tabular}{c|cc|cc}

\multirow{2}{*}{Contrast pair} & \multicolumn{2}{c|}{Model} & \multicolumn{2}{c}{Experiment} \\
 & \multicolumn{1}{c|}{($E_L$, E\_H) (keV)} & CNR & \multicolumn{1}{c|}{($E_L$, E\_H) (keV)} & CNR \\ \hline \hline
BaSO$_4$/KI & \multicolumn{1}{c|}{(33,40)} & 0.4 & \multicolumn{1}{c|}{(34,40)} & 3.1 \\ \hline
BaSO$_4$/water & \multicolumn{1}{c|}{(27,38)} & 0.6 & \multicolumn{1}{c|}{(34,39)} & 3.0 \\ \hline
KI/water & \multicolumn{1}{c|}{(25,33)} & 0.5 & \multicolumn{1}{c|}{(29,34)} & 0.9 
\end{tabular}
\caption{Comparison of the optima predicted by the model with the empirically measured ones, for the three contrast pairs of the phantom sample.\label{tab:optima}}
\end{table}
The maps of BaSO$_4$/water or KI/water CNR exhibit simpler shapes as the K-edge split the map in 3 areas:
both maxima are reached with low thresholds clearly below the K-edge and  high threshold at the K-edge value. Moreover CNR gets low values and even vanishes if both thresholds are above (upper right part) or below (bottom left part) the K-edge. 
Quantitatively, for BaSO$_4$/water contrast, the model predicts an optimum at (27 keV, 38 keV) with a CNR of 0.6 where it is (34 keV, 39 keV) with a CNR of 3.0 for the experiment.
For KI/water contrast, the model predicts an optimum at (25 keV, 33 keV) with a CNR of 0.5 where it is (29 keV, 34 keV) with a CNR of 0.9 for the experiment.
The quantitative differences between model and experiment described in that section is commented in section \ref{sec:discussion}.

\begin{figure}[h!]
\centering
\includegraphics[width=\textwidth]{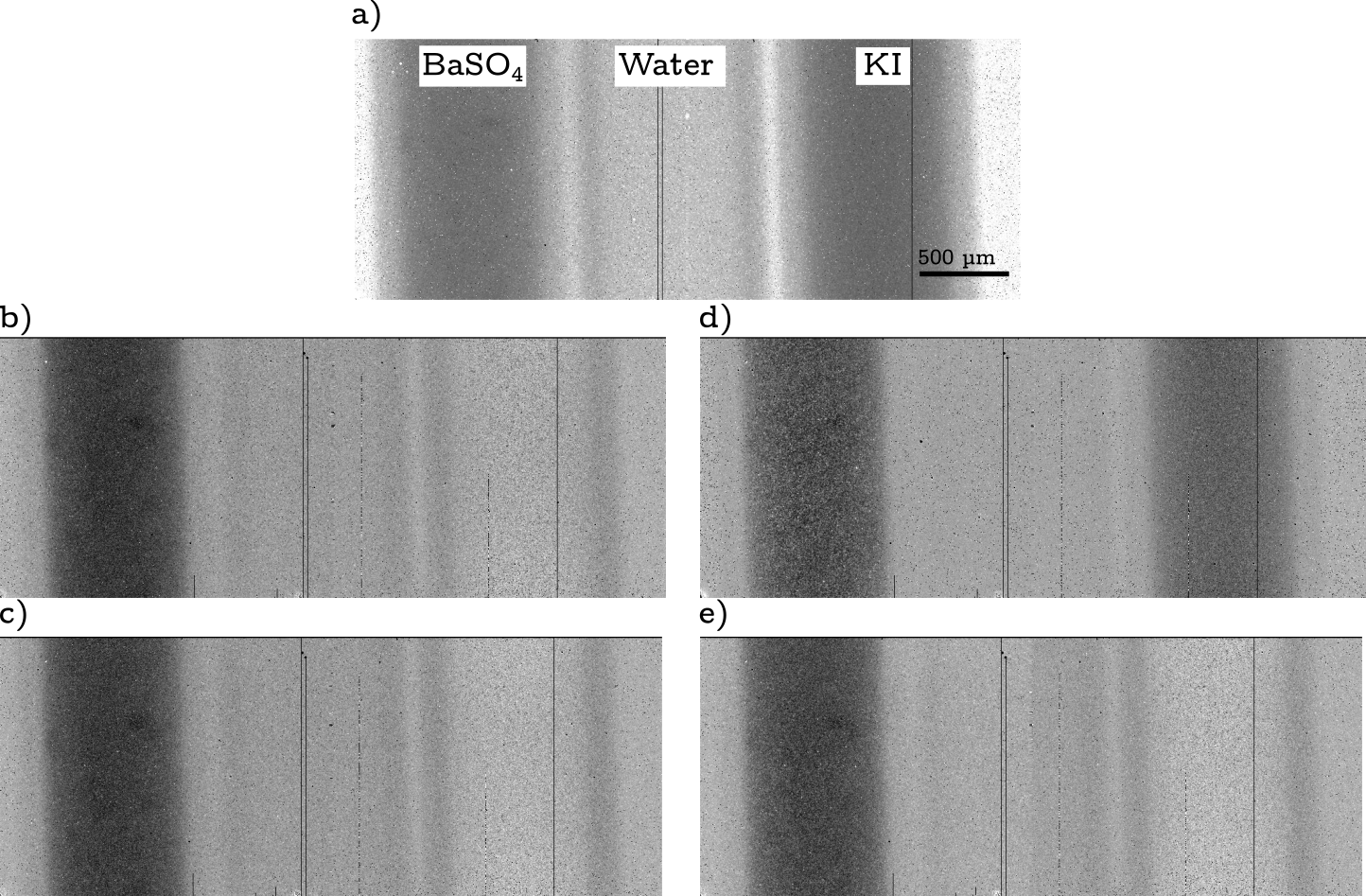}
\caption{Comparison of the phantom images  \textbf{a)}  absorption (high energy bin, $E_H=$26 keV,  \textbf{b)} KES image obtained with the modelled optima for BaSO$_4$/KI contrast ($E_L$=33 keV, $E_H$=40 keV) and \textbf{c)} KES images obtained with the empirical optima  for BaSO$_4$/KI contrast ($E_L$=34 keV, $E_H$=40 keV) \textbf{d)} KES images at modelled optima for BaSO$_4$/water contrast ($E_L$=27 keV, $E_H$=38 keV) e) KES images at empirical optima for BaSO$_4$/water contrast ($E_L$=34 keV, $E_H$=39 keV). Dark vertical lines are columns of defective pixels, ignored in the measurements of quantities related to gray level.  \label{fig:images_optima}}
\end{figure}

For a qualitative assessment of these results, figure \ref{fig:images_optima} shows images obtained a) with conventional approach, using the images from the high energy register with $E_H=26$ keV (1 phantom images and 50 flat images for flat field correction) b) KES image obtained by procedure described above at the predicted optimum for BaSO$_4$/KI contrast, i.e. $E_L$=33 keV, $E_H$=40 keV, c) KES image at empirical optimum for BaSO$_4$/KI contrast, d) KES image at the predicted optimum for BaSO$_4$/water contrast, $E_H$=38 keV and e) KES image at the empirical optimum for BaSO$_4$/water contrast. Two observations can be made: i) KES imaging permits to considerably increase the contrast between KI and BaSO$_4$, even removing water and glass capillary tubes from the image, and ii) KES images obtained at predicted and empirical optima are visually  close.

\subsection{Application example}

\begin{figure}
\centering
\includegraphics[width=0.5\textwidth]{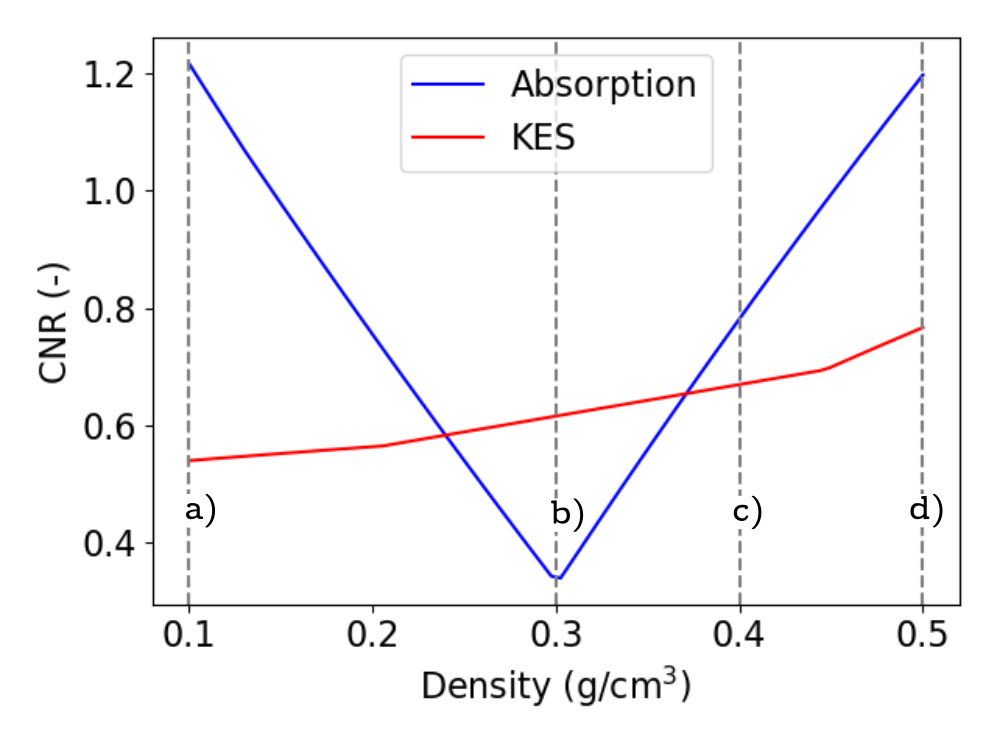}
\caption{Evolution of optimal absorption and KES CNR between BaSO$_4$  and KI as function of KI density for BaSO$_4$ density fixed at 0.46 g/cm$^3$. Letters correspond to the different panels of figure \ref{fig:kes_abs_domain}, plotting absorbance spectra at difference densities represented by the gray dashed vertical lines.\label{fig:kes_abs_domain_main}}
\end{figure}
\begin{figure}[h]
    \centering
    \includegraphics[width=0.8\textwidth]{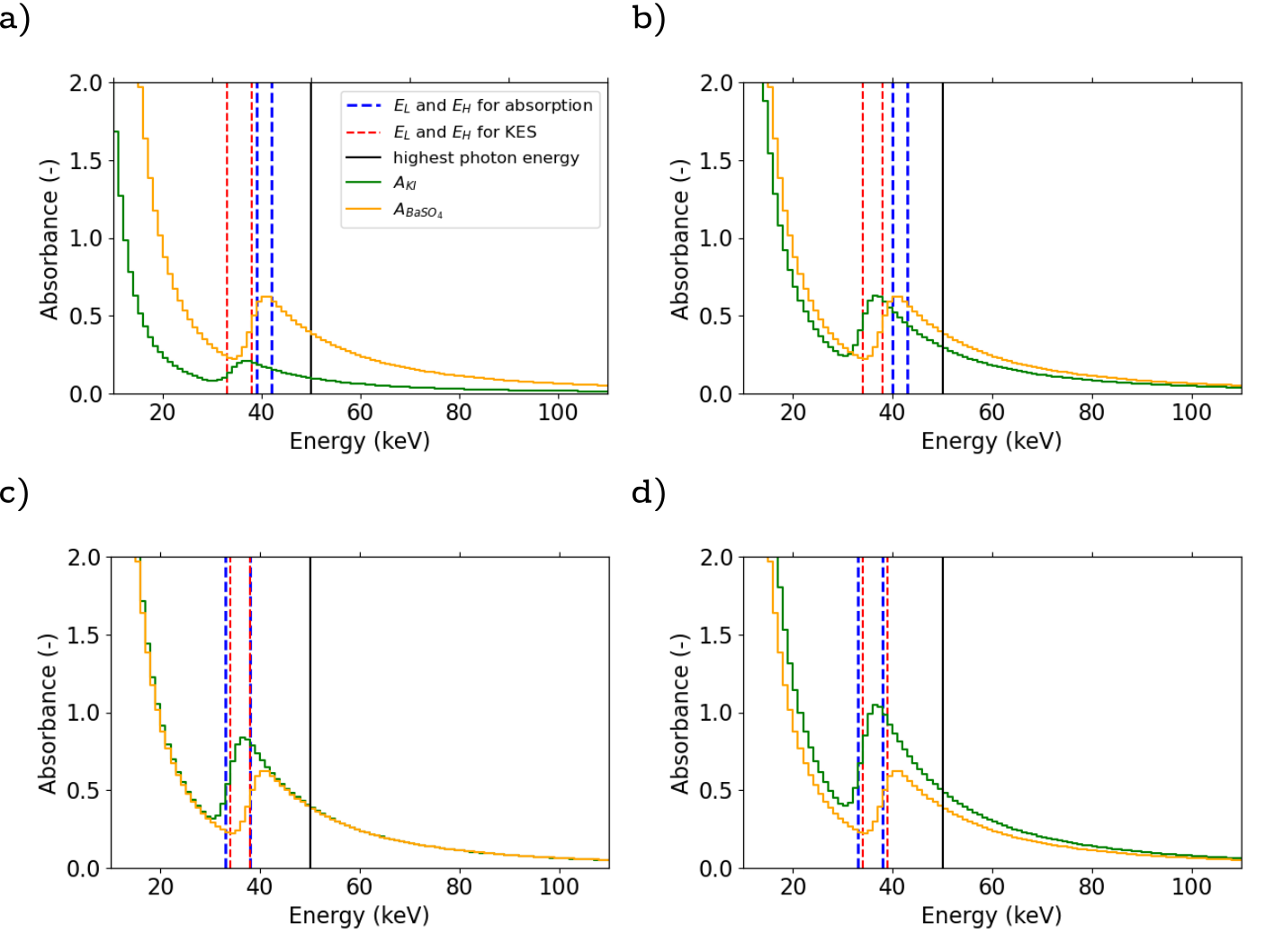}   \caption{Plot of absorbance spectra of both materials at different characteristic densities of iodine, corresponding to vertical gray dashed lines on top figure: \textbf{a)} 0.1g./cm$^3$, \textbf{b)} 0.3g/cm$^3$  \textbf{c)} 0.4 g/cm$^3$ \textbf{d)} 0.5 g/cm$^3$. Blue and red vertical dashed lines represent the optimal thresholds for absorption and KES modality respectively at the corresponding density. The vertical dark solid line represents the highest photon energy in the source spectrum with voltage at 50kV, the optimal voltage.}
    \label{fig:kes_abs_domain}
\end{figure}

Figure \ref{fig:kes_abs_domain_main} shows the absorption and KES CNR computed from equations \ref{eq:cnr_abs} and \ref{eq:cnr_kes} respectively for varying density of KI with fixed density of BaSO$_4$.
As one may see, the modality to obtain the highest CNR is not necessarily KES: it depends on the concentration of the materials. KES is recommended for density of iodine in range [0.24 g/cm$^3$, 0.37 g/cm$^3$] and absorption is recommended elsewhere.
The curve of the CNR for absorption can be split in two parts: a first decreasing and a second increasing with iodine density. Figure \ref{fig:kes_abs_domain}  a) to d)  display the optimal choice of thresholds for both modality (represented by vertical dashed lines: blue for absorption and red for KES)  in relation with the absorbance spectra of the two materials (solid curves: orange for barium, green for iodine) as well as the optimal voltage (black solid vertical line) which was found to be 50kV for all configurations. Each of these plots corresponds to different densities of iodine represented by vertical dashed line on figure \ref{fig:kes_abs_domain_main}.

\section{Discussion \label{sec:discussion}}

In section \ref{sec:noise_model}, we assumed that the dependence with $I$ and/or $E_{mean}$ can be described by $I$ only. To check that assumption, it would be interesting to reproduce the experiment by varying the source current so that $I$ varies while E$_{mean}$ does not. However, as seen in section \ref{sec:validation},  errors on $\sigma$ arise mainly from variations of E$_w$, so this assumption appears consistent with the accuracy of the model.
Also, noise level not only depends on the mean count but also on $E_W$. We suppose that this effect might be due to finite spectral resolution of the detector: Ideally, when a photon interact with the detector, it should be counted if and only if its energy lies in the energy bin defined by threshold values, i.e. the evolution of the probability for a photon to be counted presents a jump from 0 to 1 at the thresholds values. However, in practice, that probability evolves smoothly over a finite energy range centred on the thresholds values. As a result, some photons with an energy lying in that transition range may be badly counted. This results in an additional noise which increases when decreasing the gap width, as the proportion of badly classified photons increases (since the mean difference between photons energies and threshold value decreases).

Additionally, it has to be specified that the version of the PixiRad2-PixieIII that we own exhibits a salt and pepper noise when used in NPISUM mode (see figure \ref{fig:images_optima}) not mentioned in previous papers. 
This artefact is not a normal behaviour of the detector, but should be linked to errors showing up randomly in the 15bit counters of the detector.
A  first consequence of this, is that the current noise level is particularly high and globally the CNR values described above underestimate the  performances the technology has to offer.  
However, the current dependencies described by that model (on the mean count level and $E_w$ for the low energy register) are explained on the basis of other phenomenons intrinsic to the detector technology. 
Thus, we may expect that the current shape of the noise model is not affected by this artefact. 
Without that noise artefact, one may need to re-adjust the parameters of the parametrization by reproducing the exposed experiment, seeing it as a calibration procedure of the current model.

As described in section \ref{sec:validation}, figure \ref{fig:CNR_maps} a) and \ref{fig:CNR_maps} b) shows 2 maxima for the BaSO$_4$/KI pair.
Keeping in memory that KES signal measures the variation in absorption between the two energy bins, the principal maximum corresponds to the Ba K-edge as this implies strong variations of absorption for BaSO$_4$ while moderate ones for KI. 
To explain the secondary maximum it is usefull to see KES CNR as the difference of absorption CNR between the two energy bin. 
This maximum corresponds to a configuration where the high energy bin includes both K-edges so that the absorption CNR between both material on that energy bin is low. 
It remains that the KES CNR in that configuration is made only from the absorption CNR on the lower energy bin which focus on the monotonically decreasing part of the spectrum.
The minimum between these maxima corresponds to a configuration where the lower energy bin is centred on the lower K-edge and the higher energy bin on the highest K-edges.
In that configuration, the contrast vanishes : for each bin the absorption is stronger for one material on the first part of the bin, and stronger for the second material in the second part of the bin so that absorption CNR is low on both energy bin.

Finally, the application example illustrates how the proposed model allows to take into consideration different strategies and put them into balance. In that particular case, the best modality predicted by the model depends on the density of KI. 

To explain these variations, let us first focus on the absorption represented by the blue solid curve on figure \ref{fig:kes_abs_domain_main} and the blue dashed line on figure \ref{fig:kes_abs_domain}. 
For low KI density, BaSO$_4$ has the highest contrast and the difference is at its highest value above Ba K-edge; the strategy for point a) and b) is then to focus on that part of the spectrum (E$_L\simeq 40keV$) and to choose the high threshold value so to maximize the counts on both detectors (i.e. $E_H$ is set so that CL is reached on both detectors).
As the iodine density increases, the spectral contrast above the Ba K-edge reduces and gets inverted between the two K-edges (iodine becomes more absorbent than barium).
This explains why the absorption CNR decreases with iodine density up to 0.3 $g/cm^3$. 
At this point, the best strategy for absorption modality is to focus on the energy range where KI is the most absorbent.
Then, the thresholds are set so that the low energy bin focuses on the part where iodine is the most absorbent.
From here, since the CNR is built from the fact that KI is the most absorbent, the absorption CNR keeps increasing with KI density. 

Now let us focus on the KES modality illustrated by the red solid curve on figure \ref{fig:kes_abs_domain_main}, and red dashed line on figure \ref{fig:kes_abs_domain} a)-d). 
The strategy for KES is to focus the low energy bin on the part between both K-edges, and the high energy bin above the Ba K-edge. One may notice that the curve for KES CNR is much flatter than for the absorption one. This may be an interesting fact to ensure an acceptable CNR in a sample containing areas with variable density of KI or unknown composition.

The presented model allows reproducing qualitatively well the variations of CNR with energy thresholds of the detector. Particularly, the shapes of the maps on figure \ref{fig:CNR_maps} are similar to the experimental ones, and the predicted optimum is quantitatively correct for the case where KES imaging is the most relevant.
However, noticeable errors remain. 
Both flux and noise standard deviation on detector after going through the materials show increasing error for small energy bins (i.e. for $E_H$ and $E_L$ close). 
As the mean intensity is used by the noise model to estimate the noise standard deviation, it is expected that the error obtained in $\overline{I_{m}^{\Omega_i}}$ propagates in $\sigma_{m}^{\Omega_i}$ and at the end in the computed CNR.
The errors obtained on  $\overline{I_{m}^{\Omega_i}}$ originates from the input flux model $I_0^{\Omega_i}(E)$ and sample absorbance model $A_m(E)$. Additionally, the CNR  computed in section \ref{sec:validation} appeared underestimated up to a factor $\simeq 10$ for KI/BaSO$_4$ contrast. This quantitative errors may have several explanations:
\begin{itemize}
\item The exposed results demonstrate the performances when used without prior empirical knowledge of the other elements from the imaging chain, i.e. this demonstrates the minimal performance of the model. 
The results may be improved if one has i) empirical source spectra $I_0^{\Omega_i}(E)$ ii) empirical absorbance spectrum of the samples $A_m(E)$. Both these spectra may be measured using Pixie-III by performing a threshold scan or with a spectrometer.

\item As shown in section \ref{sec:noise_model}, the noise depends on both mean counts and thresholds, i.e. it shows a spectral sensibility. 
However, although the developed noise model fits well to the data for images without the sample, the terms  $\sigma_m^{\Omega_i}$  are computed taking into account the effect of the sample on the mean counts only, but not on the  spectrum shape. 
Thus, it may happen that the parametrization does not give accurate estimations for those terms, particularly when $E_w$ is small. 
This would be a point to be studied in finer details. 

\item The ratio $\frac{\overline{I_{m}^{\Omega_i}}}{ \sigma_{m}^{\Omega_i}}$ ranges from 1 to 5 in the validation example, and is about 1.2 for the coordinates of the optimum.  Additionally, the specific salt and pepper noise described at the beginning of that section presents values strongly different than the mean value. For both reasons, the  presented Taylor expansions may not hold. For example, experimentally, in eq. \ref{eq:var_ln}, the first order term (second term in the right-hand side) represents 0.17 times the zero order term, which is not completely negligible yet. 
\end{itemize}

From figure \ref{fig:CNR_maps}, particularly for BaSO$_4$ contrast, one can see that the model predicts a wider low energy bin than the one found experimentally (lower $E_L$). 
Taking into account the efficiency of the detector may improve that aspect. Indeed, at the moment, the estimation of the mean counts does not take into account detector specificities : such as efficiency and linearity. 
It is anticipated that the efficiency decreases significantly for energy above the Cd and Te K-edges (26.7 keV and 31.8 keV respectively) since part of the incoming photons leads to fluorescence. 
The bin width results from a trade-off between KES signal and statistics.
Thus, in practice, energies just below the sample K-edges are under-weighted in the measurement  of the absorbance on the low energy bin in comparison to the lowest energies of that bin. 
The absorbance measured on that bin is then larger than expected (as spectral absorbance is monotonically decreasing below the K-edge) and thus the KES signal is reduced. 
As a consequence, this pushes to take narrower low energy bin in order to compensate that loss of KES signal. The optimum is then moved towards higher $E_L$ value.
This effect should increase with the K-edge energy of the sample as the efficiency decreases, this is actually the case when comparing  KI/water optima to BaSO$_4$/water optima: the error on $E_L$ is higher for BaSO$_4$/water than for KI/water. 
Additionally, this effect is not observable for the KI/BaSO$_4$ as in that case the trade-off is not to mix KES signal from both K-edges as they would annihilate each other. 
Both model and experiment indicate to use $E_L$ just above the lower K-edge.

Finally, the optimization procedure described by algorithm \ref{algo} is quite simple. One may consider developing it by adding some constrains on the product exposure time times source  intensity. Indeed, when asking a count level CL in the image, the algorithm adapts that product to reach that level. In practice, both parameters are limited. The source is physically limited by its design and the exposure time is in practice limited to get a reasonable acquisition time or by the need in time resolution for in-situ experiment for example.  Additionally, one may include additional modalities. 
For example, it would be straightforward to consider the absorption image obtained uniquely on the low energy bin. This would consider an absorption modality where some low and high energies are ignored. This may be interesting in a case like the one depicted in figure \ref{fig:kes_abs_domain}.b where the contrast is mainly due to the absorption difference in the low energy bin.
Since the high energy image is contrasted in the opposite manner, summing the counts from both energy bins finally reduces the global contrast.

\section{Conclusions and perspectives}

We designed a model that estimates CNR of both absorption and KES modalities for Pixie III photon counting detector. The model optimizes the setup parameters : voltage U, the product exposure time times intensity, thresholds $E_L$ and $E_H$ and acquisition modality (conventional absorption or KES). The model can be divided into sub-models that can be used and improved independently. Particularly, we have presented:

\begin{itemize}
\item A noise model that evaluates the standard deviation of count levels on a representative ROI of a homogeneous material. Besides accounting for intrinsic Poisson noise which evolves as the square root of the mean intensity, the noise model includes an empirical parametrization of Pixie III  that reveals and takes into account a specific behaviour of the detector for small gap with  $E_w$.
The resulting parametrization reproduces quantitatively the data.
\item A noise propagation model that evaluates CNR for absorption and KES modalities on the projection images at the end of the processing chain.
\end{itemize}

Computationally speaking, the model  reduces to  simple operations on spectral integrals which lead to instantaneous computations.
The optimization consists in an exhaustive exploration of the space of acquisition parameters given a desired count level on the images.
The output typically consists of CNR maps in the space of acquisition parameters.
The simulated CNR maps match qualitatively and qunatitatively the experimental ones. The prediction compares within few keV to experimental values. 
The model predicts lower threshold values than the ones suggested by experiment.
Quantitatively, the error increases with small energy bins.
Additionally, it has to be kept in mind that the detector currently in our possession suffer from a random noise artefact. 
That should not impact the predicted optimal parameters, but may require recalibration of the noise model for quantitative predictions. 
Finally, it has been illustrated how the model enables to compare different strategies to enhance the contrast and how it allows understanding the underlying trade-offs the operator has to make.

To improve this model, one may consider:
\begin{itemize} 
\item to push forward the noise model by taking into account the effect of sample on the spectrum of the X-ray beam.
\item to account for linearity and efficiency of the detector. Particularly, this would improve the predictions for sample with K-edge above CdTe K-edges.
\item to design more advanced resolution algorithms, this includes : additions of derivate modalities in the comparison, constrains on the space of set up parameters, computational improvement of the optimization scheme.
\end{itemize}

\section*{Supplementary Materials}

The model was implemented in Python and is available on gitlab : \url{https://gricad-gitlab.univ-grenoble-alpes.fr/TomoX_SIMaP/px3opt}

\acknowledgments
This project was funded by the French National Research Agency ANR-18-CE42-0005. 3SR is part of LabEx Tec 21 - ANR-11-LABX-0030 and of Institut Carnot PolyNat (ANR16-CARN-0025). SIMAP is part of LabEx CEMAM (ANR-10-LABX-44-01).

\bibliographystyle{JHEP}
\bibliography{biblio}

\providecommand{\href}[2]{#2}\begingroup\raggedright\begin{thebibliography}{10}

\bibitem{maire2014}
E.~Maire and P.~J. Withers, \emph{Quantitative {{X-ray}} tomography},
  \href{http://dx.doi.org/10.1179/1743280413Y.0000000023}{\emph{International
  Materials Reviews} {\bfseries 59} (Jan., 2014) 1--43}.

\bibitem{nugent1996}
K.~A. Nugent, T.~E. Gureyev, D.~F. Cookson, D.~Paganin and Z.~Barnea,
  \emph{Quantitative phase imaging using hard x rays}, {\emph{Physical review
  letters} {\bfseries 77} (1996) 2961}.

\bibitem{paganin2006}
D.~Paganin, \emph{Coherent {{X-ray}} Optics}.
\newblock No.~6. {Oxford University Press on Demand}, 2006.

\bibitem{myers2007}
G.~R. Myers, S.~C. Mayo, T.~E. Gureyev, D.~M. Paganin and S.~W. Wilkins,
  \emph{Polychromatic cone-beam phase-contrast tomography},
  \href{http://dx.doi.org/10.1103/PhysRevA.76.045804}{\emph{Physical Review A}
  {\bfseries 76} (Oct., 2007) 045804}.

\bibitem{thomlinson2018}
W.~Thomlinson, \emph{K-edge subtraction synchrotron {{X-ray}} imaging in
  bio-medical research}, {\emph{Physica Medica} (2018) 19}.

\bibitem{jacobson1953}
B.~Jacobson, \emph{Dichromatic {{Absorption Radiography}}. {{Dichromography}}},
  \href{http://dx.doi.org/10.3109/00016925309136730}{\emph{Acta Radiologica}
  {\bfseries 39} (June, 1953) 437--452}.

\bibitem{rutt1983}
B.~K. Rutt, I.~A. Cunningham and A.~Fenster, \emph{Selective iodine imaging
  using lanthanum {{K}} fluorescence},
  \href{http://dx.doi.org/10.1118/1.595447}{\emph{Medical Physics} {\bfseries
  10} (1983) 801--808}.

\bibitem{zhong1997}
Z.~Zhong, D.~Chapman, R.~Menk, J.~Richardson, S.~Theophanis and W.~Thomlinson,
  \emph{Monochromatic energy-subtraction radiography using a rotating anode
  source and a bent {{Laue}} monochromator},
  \href{http://dx.doi.org/10.1088/0031-9155/42/9/007}{\emph{Physics in Medicine
  and Biology} {\bfseries 42} (Sept., 1997) 1751--1762}.

\bibitem{rubenstein1984}
E.~Rubenstein, \emph{Medical imaging with synchrotron radiation},
  \href{http://dx.doi.org/10.1016/0167-5087(84)90548-9}{\emph{Nuclear
  Instruments and Methods in Physics Research} {\bfseries 222} (May, 1984)
  302--307}.

\bibitem{elleaume2002}
H.~Elleaume, A.~M. Charvet, S.~Corde, F.~{Est~ve} and J.~F.~L. Bas,
  \emph{Performance of computed tomography for contrast agent concentration
  measurements with monochromatic x-ray beams: Comparison of {{K-edge}} versus
  temporal subtraction},
  \href{http://dx.doi.org/10.1088/0031-9155/47/18/307}{\emph{Physics in
  Medicine and Biology} {\bfseries 47} (Sept., 2002) 3369--3385}.

\bibitem{kulpe2018}
S.~Kulpe, M.~Dierolf, E.~Braig, B.~G{\"u}nther, K.~Achterhold, B.~Gleich
  et~al., \emph{K-edge subtraction imaging for coronary angiography with a
  compact synchrotron {{X-ray}} source},
  \href{http://dx.doi.org/10.1371/journal.pone.0208446}{\emph{PLOS ONE}
  {\bfseries 13} (Dec., 2018) e0208446}.

\bibitem{ballabriga2016}
R.~Ballabriga, J.~Alozy, M.~Campbell, E.~Frojdh, E.~H.~M. Heijne, T.~Koenig
  et~al., \emph{Review of hybrid pixel detector readout {{ASICs}} for
  spectroscopic {{X-ray}} imaging}, {\emph{Journal of Instrumentation}
  {\bfseries 11} (2016) }.

\bibitem{bellazzini2015}
R.~Bellazzini, A.~Brez, G.~Spandre, M.~Minuti, M.~Pinchera, P.~Delogu et~al.,
  \emph{{{PIXIE III}}: A very large area photon-counting {{CMOS}} pixel
  {{ASIC}} for sharp {{X-ray}} spectral imaging},
  \href{http://dx.doi.org/10.1088/1748-0221/10/01/C01032}{\emph{Journal of
  Instrumentation} {\bfseries 10} (Jan., 2015) C01032--C01032}.

\bibitem{he2012}
P.~He, B.~Wei, W.~Cong and G.~Wang, \emph{Optimization of {{K-edge}} imaging
  with spectral {{CT}}},
  \href{http://dx.doi.org/10.1118/1.4754587}{\emph{Medical Physics} {\bfseries
  39} (Nov., 2012) 6572--6579}.

\bibitem{he2013}
P.~He, B.~Wei, P.~Feng, M.~Chen and D.~Mi, \emph{Material {{Discrimination
  Based}} on {{K-edge Characteristics}}},
  \href{http://dx.doi.org/10.1155/2013/308520}{\emph{Computational and
  Mathematical Methods in Medicine} {\bfseries 2013} (2013) 308520}.

\bibitem{meng2016}
B.~Meng, W.~Cong, Y.~Xi, B.~De~Man and G.~Wang, \emph{Energy {{Window
  Optimization}} for {{X-ray K-edge Tomographic Imaging}}},
  \href{http://dx.doi.org/10.1109/TBME.2015.2413816}{\emph{IEEE transactions on
  bio-medical engineering} {\bfseries 63} (Aug., 2016) 1623--1630}.

\bibitem{brun2020}
F.~Brun, V.~D. Trapani, J.~Albers, P.~Sacco, D.~Dreossi, L.~Brombal et~al.,
  \emph{Single-shot {{K-edge}} subtraction x-ray discrete computed tomography
  with a polychromatic source and the {{Pixie-III}} detector},
  \href{http://dx.doi.org/10.1088/1361-6560/ab7105}{\emph{Physics in Medicine
  \& Biology} {\bfseries 65} (Mar., 2020) 055016}.

\bibitem{schoonjans2011}
T.~Schoonjans, A.~Brunetti, B.~Golosio, M.~S. {del Rio}, V.~A. Sol{\'e},
  C.~Ferrero et~al., \emph{The xraylib library for {{X-ray}}\textendash matter
  interactions. {{Recent}} developments}, {\emph{Spectrochimica Acta Part B:
  Atomic Spectroscopy} {\bfseries 66} (2011) 776--784}.

\bibitem{benaroya2005}
H.~Benaroya, S.~M. Han and M.~Nagurka, \emph{Probability {{Models}} in
  {{Engineering}} and {{Science}}}.
\newblock {CRC Press}, June, 2005.

\bibitem{ditrapani2018}
V.~Di~Trapani, L.~Brombal, S.~Donato, B.~Golosio, R.~Longo, P.~Oliva et~al.,
  \emph{36. {{Optimization}} of the acquisition threshold of {{Photon Counting
  Detectors}} ({{PCDs}}) used in {{X-ray}} medical imaging},
  \href{http://dx.doi.org/10.1016/j.ejmp.2018.04.046}{\emph{Physica Medica}
  {\bfseries 56} (Dec., 2018) 84}.

\bibitem{ditrapani2020}
V.~Di~Trapani, A.~Bravin, F.~Brun, D.~Dreossi, R.~Longo, A.~Mittone et~al.,
  \emph{Characterization of the acquisition modes implemented in
  {{Pixirad-1}}/{{Pixie-III X-ray Detector}}: {{Effects}} of charge sharing
  correction on spectral resolution and image quality},
  \href{http://dx.doi.org/10.1016/j.nima.2019.163220}{\emph{Nuclear Instruments
  and Methods in Physics Research Section A: Accelerators, Spectrometers,
  Detectors and Associated Equipment} {\bfseries 955} (Mar., 2020) 163220}.

\bibitem{poludniowski2021}
G.~Poludniowski, A.~Omar, R.~Bujila and P.~Andreo, \emph{Technical {{Note}}:
  {{SpekPy}} v2.0\textemdash a software toolkit for modeling x-ray tube
  spectra}, \href{http://dx.doi.org/10.1002/mp.14945}{\emph{Medical Physics}
  {\bfseries 48} (2021) 3630--3637}.

\bibitem{bujila2020}
R.~Bujila, A.~Omar and G.~Poludniowski, \emph{A validation of {{SpekPy}}: {{A}}
  software toolkit for modelling {{X-ray}} tube spectra},
  \href{http://dx.doi.org/10.1016/j.ejmp.2020.04.026}{\emph{Physica Medica}
  {\bfseries 75} (July, 2020) 44--54}.

\end{thebibliography}\endgroup
\end{document}